# Comprehensive Demonstration of Spin-Hall Hanle Effects in Epitaxial Pt Thin Films

Jing Li[1*], Andrew H. Comstock[2*], Dali Sun[2⊥], and Xiaoshan Xu[1⊥]

[1]Department of Physics and Astronomy and Nebraska Center for Materials and Nanoscience, University of Nebraska, Lincoln, NE 68588, USA

[2]Department of Physics and Organic and Carbon Electronics Lab (ORaCEL), North Carolina State University, Raleigh, NC 27695, USA

* Authors with equal contributions

⊥ xiaoshan.xu@unl.edu, dsun4@ncsu.edu

**Abstract**

We demonstrate a nonlinear Hall effect due to the boundary spin accumulation in Pt films grown on $Al_2O_3$ substrates. This Hall effect and the previously demonstrated Hanle magnetoresistance provide a complete picture of the spin-precession control of the spin and charge transport at the boundary of a spin-orbit coupled material, which we refer to as spin-Hall Hanle effects (SHHE). We also show that the SHHE can be employed to measure the spin diffusion length, the spin-Hall angle, and the spin relaxation time of heavy metal without the need of magnetic interface or the input from other measurements. The comprehensive demonstration of SHHE in such a simple system suggests they may be ubiquitous and needs to be considered for unravelling the spin and charge transport in more complex thin film structures of spin-orbit coupled materials.

Ever since the discovery of the spin Hall effect (SHE) and the inverse spin Hall effect (ISHE) [1–5], the strong spin-orbit coupling in materials such as heavy metals has been widely used for both the generation [6] and detection of pure spin current [7]. Recently, the interaction between spin current (from the SHE) in a heavy metal and the local spins in an adjacent magnetic material has been demonstrated to give rise to the spin-Hall magnetoresistance (SMR) [8,9]. This interfacial mechanism was later employed to effectively detect [10–12] and even manipulate [13] the antiferromagnetic order which is beyond the capability of bulk magnetometry.

On the other hand, it is difficult to probe SHE and ISHE as bulk effects in heavy metals using magneto-transport, because the modulation by a magnetic field is limited by the short timescale of momentum relaxation [14]. In contrast, at the boundary of the heavy metals, the spin accumulation and diffusion (spin-current reflection) may be more effectively manipulated by a magnetic field via spin precession according to the Hanle effect [Fig. 1(a)], because it is the spin relaxation that determines the timescale of the process. Given the relationship between spin diffusion and charge current as described by the ISHE, Dyakonov predicted a longitudinal effect [15], which was later observed in Pt and β-phase Ta thin films [16,17] and named Hanle magnetoresistance; one key evidence is the anisotropy since spin precession depends on the angle between the magnetic field and the initial spin polarization [16].

What's puzzling is the transverse effect. In principle, spin precession is expected to rotate the spin polarization and generate a transverse charge current corresponding to a Hall effect. However, this Hall effect has not been experimentally demonstrated and often overlooked. In particular, in the previous work [16] where the longitudinal effect was demonstrated, only a linear field-dependence of the transverse signal was observed and attributed to the ordinary Hall effect (OHE).

To resolve the puzzle of missing transverse effect, we note that the previous work [16] may have only probed the weak-precession condition due to the short spin relaxation time $\tau_s$. In the weak-precession condition, a linear field dependence of the Hall effect is expected, which cannot be distinguished from the linear OHE background; meanwhile a quadratic field dependence of the magnetoresistance is expected which is consistent with the observation [16].

In this work, we perform a magneto-transport study on the interplay of SHE, ISHE, spin diffusion, and spin relaxation in Pt thin films deposited epitaxially on $Al_2O_3$ substrates using pulsed laser deposition to enhance $\tau_s$ for reaching the strong-precession condition. We observed non-quadratic and non-linear field dependence for the longitudinal (magnetoresistance) and the transverse (Hall) effects respectively, indicating the strong precession condition. The dual effects, which we refer to as the spin-Hall Hanle effects (SHHE), can be fit using the same set of parameters (spin Hall angle $\theta_{SH}$, spin diffusion length $\lambda_s$, spin relaxation time $\tau_s$), suggesting that SHHE can be reliably employed in extracting the spin transport properties without complications from the magnetic interfaces, such as spin memory loss [18] and proximity-induced magnetism [19,20].

Pt (111) thin films of various thickness were epitaxially grown on $Al_2O_3$ (0001) substrates by pulsed laser deposition with a YAG laser (266 nm wavelength, 70 mJ pulse energy, 3 Hz repetition rate) in $10^{-7}$ torr vacuum at room temperature and subsequently patterned into Hall bars by photolithography and ion milling. Crystal orientation of the Pt (111) films was confirmed using X-ray diffraction
(Sec. S1 within the Supplemental Material [21]) while the film thickness was measured using X-ray reflectivity. Longitudinal and transverse resistivity was measured using the Hall bar (Sec. S2 within the Supplemental Material [21]) in magnetic field along different directions at room temperature; the field dependence of the longitudinal ($\rho_L$) and transverse ($\rho_T$) resistivity was symmetrized and antisymmetrized respectively to minimize the spurious effects from imperfect device geometry.

Figure 1(b) shows the change of longitudinal resistivity $\Delta\rho_L=\rho_L-\rho_{L0}$ normalized with respect to the zero-field value $\rho_{L0}$ in a 5.2-nm-thick Pt film, where $B_x$, $B_y$ and $B_z$ represent the magnetic field applied along the $x$, $y$, and $z$ direction respectively. Overall, $\Delta\rho_L$ increases with the magnetic field, consistent with the expectation from the SHHE [15]. As illustrated in Fig. 1(a), the longitudinal charge current ($\vec{J}_c \parallel +\hat{x}$) in the Pt film generates a spin current ($\vec{J}_s \parallel -\hat{z}$) via SHE toward the Pt/Al$_2$O$_3$ interface with spin polarization $\vec{s} \parallel -\hat{y}$. The reflected spin current ($\vec{J}_{S,R} \parallel +\hat{z}$) generates a longitudinal charge current $\vec{J}_{C,R} \propto \vec{J}_{S,R} \times \vec{s} \parallel +\hat{x}$ via the ISHE before the spin polarization relaxes, resulting in an overall reduction in the resistivity of the Pt film. The Hanle effect may be observed when an external magnetic field causes the precession of the spin polarization of $\vec{J}_{S,R}$. In this case, the projection of $\vec{J}_{C,R}$ on $+\hat{x}$ will be reduced, which increases the longitudinal resistivity, as observed in Fig. 1(b) consistent with that in previous work [16].

The anisotropy in Fig. 1(b) also agrees with SHHE in that $\Delta\rho_L(B_y)/\rho_{L0}$ is smaller than $\Delta\rho_L(B_x)/\rho_{L0}$ and $\Delta\rho_L(B_z)/\rho_{L0}$ while the latter two are similar. When the magnetic field is parallel to the initial polarization direction ($\hat{y}$) of $\vec{J}_{s,R}$, no spin precession is caused by the external magnetic field and the SHHE does not contribute to $\Delta\rho_L(B_y)/\rho_{L0}$. As a result, ordinary magnetoresistance (OMR) is responsible for the non-zero $\Delta\rho_L(B_y)/\rho_{L0}$ observed in Fig. 1(b) which is proportional to $B^2$; hence the difference $\Delta\rho_L(B_z)-\Delta\rho_L(B_y)$ is attributed to the longitudinal SHHE.

Fig. 1(b) also reveals the strong-precession behavior of the longitudinal SHHE that was not observed before. Considering both the film-substrate and the film-vacuum boundaries, SHHE with $B_z$ can be described using [15] (Sec. S3 within the Supplemental Material [21]):

$$\frac{\Delta\rho_{SHHE}}{\rho_{L0}} = \theta_{SH}^2 \frac{\tanh(d/2\lambda_s)}{d/2\lambda_s}\left[1 - \frac{\tanh\left(\frac{\kappa d}{2\lambda_s}\right)}{\kappa \cdot \tanh\left(\frac{d}{2\lambda_s}\right)}\right] \quad (1),$$

where the real and imaginary parts of $\Delta\rho_{SHHE}$ are the longitudinal $\Delta\rho_{L,SHHE}$ and the transverse $\rho_{T,SHHE}$ respectively, $d$ is the film thickness, $\kappa = (1 - i\Omega\tau_s)^{1/2}$ is a complex quantity with $i=\sqrt{-1}$, $\Omega = g\mu_B B_z/\hbar$ is the Larmor frequency with $g$ the gyromagnetic factor, $\mu_B$ the Bohr magneton, and $\hbar$ the reduced Planck constant. A numeric simulation is displayed in Fig. 2. According to Eq. (1) and Fig. 2(a), at low field (weak precession), the longitudinal SHHE is quadratic ($\propto B_z^2$) as observed previously [16]; at high field (strong precession), the effect saturates when the precession angle is so large that the projection of $\vec{J}_{cR}$ on $\hat{x}$ cancels, consistent with the reduced slope $\Delta\rho_L(B_x)/\rho_{L0}$ and $\Delta\rho_L(B_z)/\rho_{L0}$ at high field in Fig. 1(b).

Figure 1(c) shows the normalized transverse resistivity $\rho_T/\rho_{L0}$, which has non-trivial field dependence only in $B_z$. In addition, $\rho_T(B_z)/\rho_{L0}$ exhibits a non-linear relation with a large slope at low field and a smaller slope at high field. The latter is expected to come from the ordinary Hall effect (OHE) in a non-magnetic metal. Similar field dependence of the transverse resistivity has been observed in Pt/ferrimagnetic insulator (FMI) systems, which was explained as anomalous Hall effect caused by magnetic proximity [19,22]. Here we don't have the complications from the magnetic order of the substrate, so the non-linear part of the transverse signal can be directly ascribed to SHHE as explained in the following.

As illustrated in Fig. 1(a), with $B_z$, the spin precession leads to non-zero projection of the spin polarization of $\vec{J}_{S,R}$ on $\hat{x}$, which generates a non-zero projection of $\vec{J}_{C,R}$ on $\hat{y}$ (Hall signal) via ISHE. At low field (weak precession), the effect is linear ($\propto B_z$) according to Eq. (1) and Fig. 2(b). At high field (strong precession), the transverse effect is expected to vanish because the projection of $\vec{J}_{C,R}$ on $\hat{y}$ cancels due to the large precession angle. This overall nonlinear effect is consistent with the observation in Fig. 1(c).

The observation of the non-quadratic longitudinal and the non-linear transverse field dependence in Figs. 1(b) and (c) respectively, suggests the strong-precession condition of SHHE in Eq. (1). In principle, all the parameters contributing to SHHE, i.e., $\tau_s$, $\theta_{SH}$, and $\lambda_s$ can be extracted by fitting the experimental data using the field dependence in Eq. (1). On the other hand, a scaling rule pointed out by Dyakonov [15] (Sec. S3 within the Supplemental Material [21]) also needs to be considered, as described below.

Considering the spin-precession nature, the SHHEs are expected to scale with the spin precession time $\tau_s^*$. For $d/\lambda_s \to \infty$ (thick film limit), $\tau_s^*$ is limited by the spin relaxation time $\tau_s$, i.e., $\tau_s^* = \tau_s$, as illustrated in Fig. 1(a). For $d/\lambda_s \to 0$ (thin film limit), spin precession occurs over the entire film thickness, so $\tau_s^*$ is the same as the spin diffusion time $\tau_D = \frac{\left(\frac{d}{2}\right)^2}{D} = \tau_s \left(\frac{d}{2\lambda_s}\right)^2$, where $D = \frac{\lambda_s^2}{\tau_s}$ is the spin diffusion coefficient. Dyakonov then introduced the definition $\frac{1}{\tau_s^*} = \frac{1}{\tau_s} + \frac{1}{\tau_D} = \frac{1}{\tau_s}\left[1 + \left(\frac{2\lambda_s}{d}\right)^2\right]$ to describe the dependence of $\tau_s^*$ on both $\tau_s$ and $d$ [15]. As shown in Fig. 2, $\Delta\rho_{L,SHHE}$ and $\rho_{T,SHHE}$ simulated according to Eq. (1) are normalized with the maximum longitudinal effect $\Delta\rho_{L,SHHE}(B_z=\infty)$ and plotted as a function of $\Omega\tau_s^*$. Indeed, the "scaled" field dependence of SHHE maintains roughly the same curve shape despite that the value of $d/\lambda_s$ changes over four orders of magnitude.

The Dyakonov's scaling rule suggests that it is difficult to unambiguously determine $\tau_s$, $\theta_{SH}$, and $\lambda_s$ altogether by fitting the measured field dependence of $\Delta\rho_{SHHE}/\rho_{L0}$ using Eq. (1) considering the experimental uncertainty, because it is $\tau_s^*$ instead of $\tau_s$ that can be directly extracted. On the other hand, here we notice that $\lambda_s$ can be estimated based on the thickness dependence of SHHE, which can then be used to extract $\tau_s$ (out of $\tau_s^*$) and $\theta_{SH}$. A close look at Eq. (1) reveals that the low-field $\Delta\rho_{SHHE}/\rho_{L0}$ has a maximum at an intermediate film thickness because it vanishes in both the thin and thick film limits: For $d/\lambda_s \to 0$ (thin film limit), $\Delta\rho_{SHHE}$ approaches zero because $\tau_s^* \to 0$ means no precession; for $d/\lambda_s \to \infty$ (thick limit), $\Delta\rho_{SHHE}/\rho_{L0}$ also approaches zero because the effect of the spin precession that occurs at the boundary is unimportant for thick films. It turns out that the thickness for reaching maximum low-field $\Delta\rho_{SHHE}/\rho_{L0}$ only depends on $\lambda_s$, or $d/\lambda_s \approx 4.56$ and $d/\lambda_s \approx 3.28$ for $\Delta\rho_{L,SHHE}/\rho_{L0}$ and $\rho_{T,SHHE}/\rho_{L0}$ respectively (Sec. S3 within the Supplemental Material [21]), as also given by Eq. S36 and Eq. S38 in ref. [16].

Considering this property, we measured the thickness dependence of SHHE in the epitaxial Pt films. The experimental $\Delta\rho_{L,SHHE}/\rho_{L0}$ is calculated by subtracting the OMR contribution, i.e., $[\Delta\rho_L(B_z) - \Delta\rho_L(B_y)]/\rho_{L0}$. Fig. 3 shows the thickness dependence of experimental $\Delta\rho_{L,SHHE}/\rho_{L0}$ at 4 T field. Meanwhile, the experimental $\Delta\rho_{T,SHHE}/\rho_{L0}$ is calculated by subtracting the linear OHE contribution from $\rho_T(B_z)/\rho_{L0}$; the result at 1 T field is displayed in Fig. 3. Fitting the thickness dependence of both longitudinal and transverse SHHE leads to $\lambda_s=1.63 \pm 0.26$ nm. The $\lambda_s$ values are comparable to the value reported in polycrystalline Pt/sapphire at 300 K [16] and single crystalline Pt/Fe/MgO [23,24].

With the estimation of $\lambda_s$, we may fit the field dependence of SHHE signals using Eq. (1) and derive the value of $\theta_{SH}$, $\tau_s$, and the related diffusion coefficient $D=\lambda_s^2/\tau_s$. Fig. 4 shows fittings of both longitudinal and transverse SHHE signals from three different Pt/Al$_2$O$_3$ films. For each film, same set of parameters ($\theta_{SH}$, $\lambda_s$, $\tau_s$) have been used to fit both longitudinal and transverse SHHE (Sec. S4 within the Supplemental Material [21]). The derived spin transport properties of Pt are summarized in Table 1 and compared with those from Ref. [16]. One salient difference between this work and previous work [16] is that the spin relaxation time $\tau_s$ is roughly one order of magnitude longer in the epitaxial Pt films used in this work, which is critical for reaching the strong-precession condition of SHHE.

As pointed out by Dyakonov [15], for $d/\lambda_s \to 0$, the maximum longitudinal SHHE, i.e., $\Delta\rho_{L,SHHE}(\infty)/\rho_{L0}$ approaches $\theta_{SH}^2$. As a result, $\Delta\rho_{L,SHHE}(B_z=\infty)/\rho_{L0}$ measured from thin Pt films generally provides a more precise estimation of $\theta_{SH}$. Meanwhile, in thick Pt films, the spin precession time $\tau_s^*$ that determines the shape of the field dependence of SHHE is simply $\tau_s$, hence measurements from thicker Pt films generally provide a more precise estimation of $\tau_s$. Based on these arguments, we found that $\theta_{SH}$ and $\tau_s$ are most likely to be $0.022\pm0.006$ and $1.8\pm0.9$ ps, respectively in our Pt thin films.

The $\theta_{SH}$ value of our Pt films is lower than the values of $0.048\pm0.015$ [23] and $0.057\pm0.014$ [24] reported for single crystalline (001) Pt/Fe/MgO measured using spin pumping, but still lies within the range between 0.01 and 0.1 reported for polycrystalline Pt [25,26]. Crystal orientation might be responsible for the discrepancy of $\theta_{SH}$ values among different single-crystalline Pt films. It has been shown that $\theta_{SH}$ of Pt can be tuned from 1% to 10% by varying the resistivity of polycrystalline Pt films [26]. The $\theta_{SH}$ value of our Pt films is comparable to that of e-beam evaporated polycrystalline Pt films, while the longitudinal resistivity (20~50 μΩ·cm) of our Pt films at 300 K is slightly larger than that (~18 μΩ·cm) of evaporated Pt in super-clean metal regime [26]. Considering that the grain size of our (111) Pt films is small (~ 3 nm derived from x-ray diffraction), it is reasonable to postulate that abundant grain boundaries exist along charge current flow direction within Pt film plane while there are far fewer grain boundaries hindering spin current flow along the normal direction of thin film plane, which may explain the similarity of $\theta_{SH}$ values between epitaxial and polycrystalline Pt films. It is noteworthy that our epitaxial Pt films do not form interfaces with any magnetic substrates, eliminating the intricacy of separating spurious contributions, such as spin rectification effect [23,24], from ISHE contribution to the measured signals.

In conclusion, this work has demonstrated that SHHE emerges as non-linear Hall effect and non-quadratic magnetoresistance in epitaxial Pt films on Al$_2$O$_3$ substrates at room temperature. Importantly, we show that SHHE can be employed to reliably measure spin transport properties of spin-orbit coupled materials, without the complication of magnetic interfaces or the need of other measurements. The simplicity of SHHE suggests that with a magnitude up to $\theta_{SH}^2$, they are expected to be ubiquitous in heavy metal thin film systems. Recognition of the contribution of SHHE in more complex systems (e.g., with magnetic interface) can be pivotal for understanding their entangled magnetoresistance and Hall effects.

**Acknowledgment**

This research was primarily supported by the U.S. Department of Energy (DOE), Office of Science, Basic Energy Sciences (BES), under Award No. DE-SC0019173. The work at NC State was supported by the U.S. Department of Energy (DOE), Office of Science, Basic Energy Sciences (BES), under Award No. DE-SC0020992. The research was performed in part in the Nebraska Nanoscale Facility: National Nanotechnology Coordinated Infrastructure and the Nebraska Center for Materials and Nanoscience (and/or NERCF), which are supported by the National Science Foundation under Award ECCS: 1542182, and the Nebraska Research Initiative.

| Structure | Pt thickness $d$ (nm) | Spin Hall angle $\theta_{SH}$ | Spin diffusion length $\lambda_S$ (nm) | Diffusion coefficient $D$ (mm$^2$/s) | Spin relaxation time $\tau_S$ (ps) | $T$ (K) |
|---|---|---|---|---|---|---|
| Pt/Al$_2$O$_3$ | 3.8 | 0.022±0.006 | 1.82±0.07 | 1.0±0.4 | 3.9±1.9 | 300 |
| Pt/Al$_2$O$_3$ | 5.2 | 0.029±0.004 | 1.72±0.10 | 1.1±0.5 | 2.9±0.8 | 300 |
| Pt/Al$_2$O$_3$ | 6.3 | 0.040±0.015 | 1.66±0.15 | 1.9±0.8 | 1.8±0.9 | 300 |
| Pt/Pyrex [16] | 3 | 0.056 | 0.8 | 3.4 | 0.19 | 100 |
| Pt/SiO$_2$ [16] | 3 | 0.056 | 1.4 | 18 | 0.11 | 100 |

Table 1. Comparison between spin transport properties of Pt in this work and those in reference.

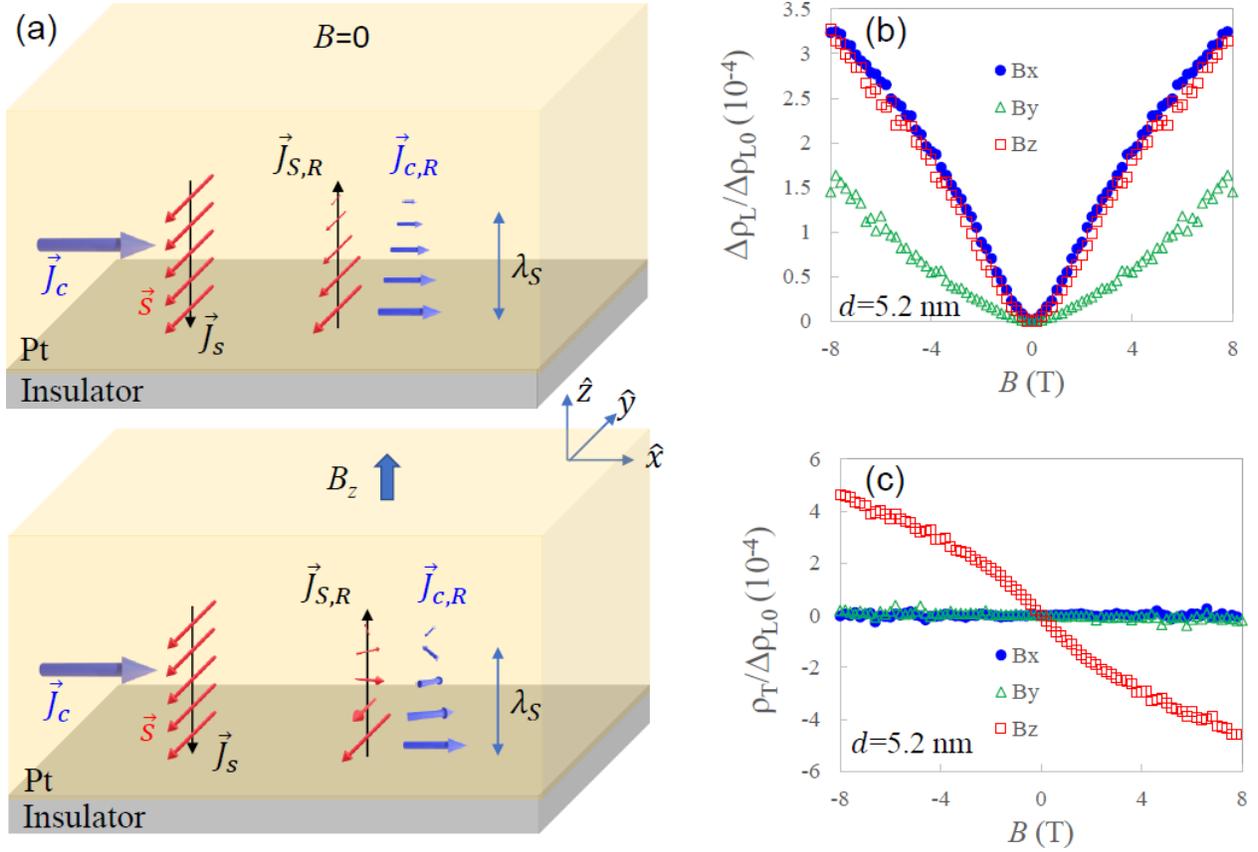

**Figure 1**. (a) Schematics of the spin-Hall Hanle effects at the film-substrate interface. Top: no spin precession in zero field. Bottom: spin precession when the field is perpendicular to the interface (z direction). $\vec{J}_c$ and $\vec{J}_s$ are the charge current and corresponding spin current generated via SHE respectively. $\vec{J}_{S,R}$ and $\vec{J}_{c,R}$ are the reflected spin current and the corresponding charge current generated via ISHE respectively. (b) Measured change of longitudinal resistivity normalized with the zero-field resistivity. $B_x$, $B_y$, and $B_z$ are the magnetic field along the $x$, $y$, and $z$ direction respectively. (c) Measured transverse Hall resistivity normalized with the zero-field longitudinal resistivity.

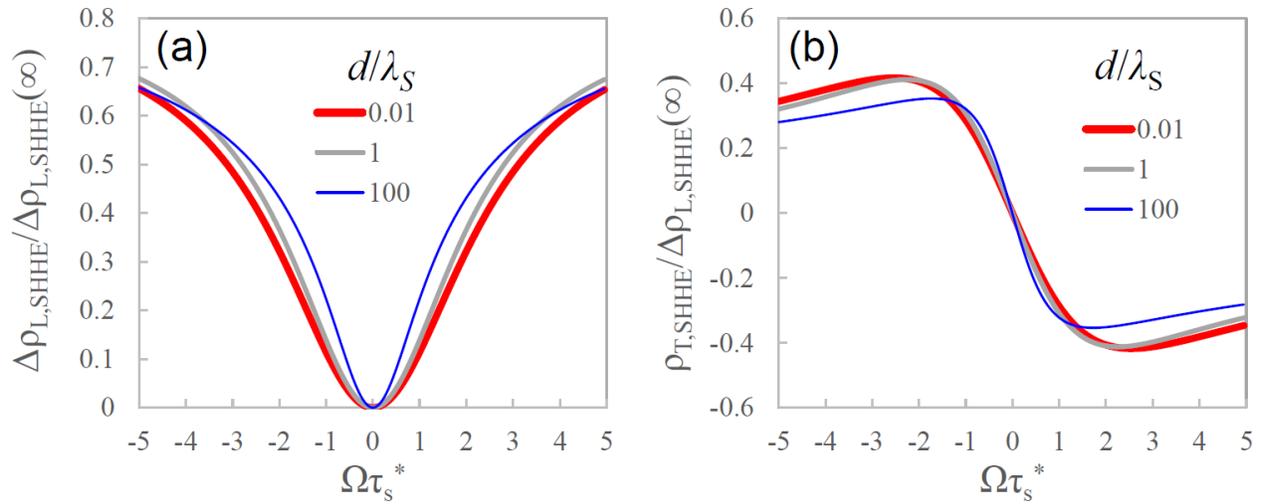

**Figure 2**. Simulated longitudinal (a) and transverse (b) SHHE for different $d/\lambda_S$ ratio according to the real and imaginary part of Eq. (1) respectively. Both the longitudinal and the transverse effects are normalized with the maximum longitudinal SHHE which is the value at infinity field. $\tau_s^*$ is the spin precession time (see text).

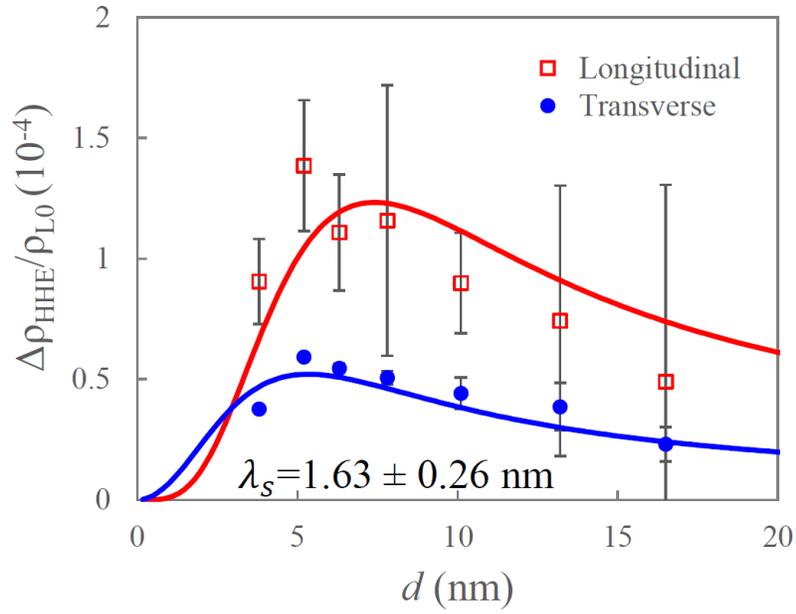

**Figure 3**. Longitudinal and transverse SHHE measured at 4 T and 1 T respectively, as a function of Pt thickness. The spin diffusion length estimated from the longitudinal and transverse effects are $1.63 \pm 0.26$ nm.

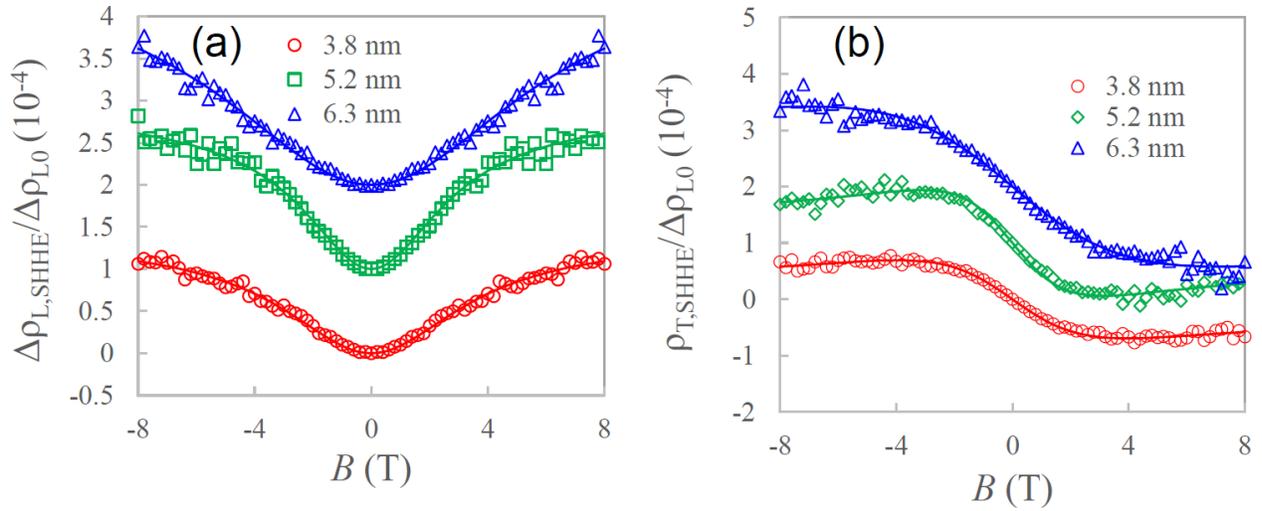

**Figure 4.** (a) Measured longitudinal SHHE (symbols) as the difference between the magnetoresistance in $B_z$ and that in $B_y$. (b) Measured transverse SHHE (symbols) as the measured Hall effect with the linear OHE background subtracted. The data are shifted vertically in both (a) and (b) for clarity. The lines in (a) and (b) are fit of the data using Eq. (1). For each Pt film thickness, the fit uses the same set of parameters.

# Comprehensive Demonstration of Spin-Hall Hanle Effects in Epitaxial Pt Thin Films

Jing Li*, Andrew H. Comstock*, Dali Sun⊥, and Xiaoshan Xu⊥

Department of Physics and Astronomy, University of Nebraska, Lincoln, NE 68588, USA

Department of Physics and Organic and Carbon Electronics Lab (ORaCEL), North Carolina State University, Raleigh, NC 27695, USA

*Authors with equal contributions

⊥ xiaoshan.xu@unl.edu , dsun4@ncsu.edu

## S1. X-ray diffraction

High-resolution x-ray diffraction (XRD) and x-ray reflectivity of epitaxial Pt films were measured using Rigaku SmartLab Diffractometer with Cu K$\alpha$ radiation (wavelength 1.54 Angstrom). As shown in Fig. S1, the Pt (111) and (222) peaks in this specular diffraction indicate that [111] is the preferential out-of-plane orientation of epitaxial Pt films grown on (0001) $Al_2O_3$ substrates.

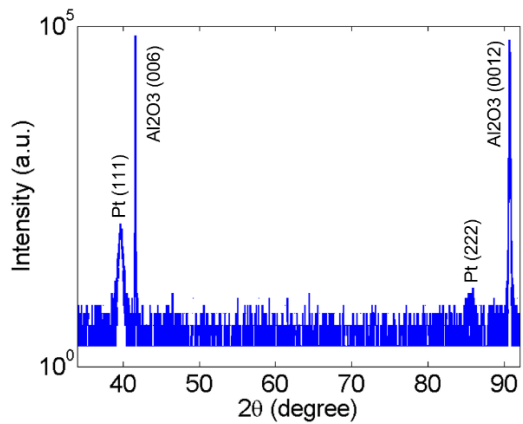

**Fig. S1** X-ray diffraction pattern of a 40-nm Pt/$Al_2O_3$ sample deposited under the same conditions as the samples used in the main text.

## S2. Magneto-transport measurement

Magneto-transport measurements were performed in a Quantum Design physical property measurement system equipped with a 9 Tesla longitudinal magnet and horizontal rotator options. The longitudinal resistivity and transverse Hall signals from a Pt Hall bar were measured using an AC Transport option with 100 μA excitation current at 17 Hz.

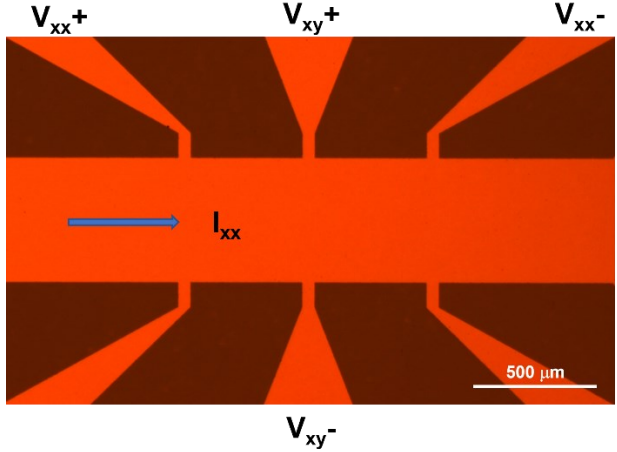

**Fig. S2** Image of a Pt Hall bar used for magneto-transport measurements.

### S3. Derivation of the spin-Hall Hanle effects (SHHEs)

**3.1 Field dependence of the longitudinal (magnetoresistance) and the transverse (Hall) effects**

We start with the equation of motion that includes the spin-Hall and inverse spin-Hall effects

$$q_i = -\sigma E_i + \theta_{SH}\epsilon_{ijk}q_{jk}$$

$$q_{ij} = -D\frac{\partial P_j}{\partial x_i} - \theta_{SH}\epsilon_{ijk}q_k$$

where $\overleftrightarrow{q}$ and $\vec{q}$ are the spin and charge currents density in the unit of charge current, $\vec{E}$ is the electric field, $\vec{P}$ is the spin polarization, $\theta_{SH}$ is the spin-Hall angle, $\epsilon_{ijk}$ is the anti-symmetric Levi-Civita symbol, $\sigma$ is the charge conductivity, $D$ is the diffusion coefficient.

The system we consider is a slab/film structure where $x$, and $y$ dimensions are much larger than the $z$ dimension ($-d/2 \leq z \leq d/2$). Assuming $B_x = B_y = 0$, $E_y = E_z = 0$, there should be no spatial variation along the $x$ and $y$ directions. Thus, $\frac{\partial P_i}{\partial x} = \frac{\partial P_i}{\partial y} = 0$.

Near the boundary ($z = \pm d/2$), the variation along the $z$ direction is governed by the continuity equation:

$$\frac{\partial P_j}{\partial t} + g\frac{\mu_B}{\hbar}(\vec{P} \times \vec{B})_j + \frac{\partial q_{ij}}{\partial x_i} + \frac{P_j}{\tau_s} = 0,$$

with the boundary condition of no spin current $q_{zj}(z = -\frac{L}{2},\frac{L}{2}) = 0$, where $g$ is the Lande g-factor, $\mu_B$ is the Bohr magneton, $\hbar$ is the reduced Plank constant, $\tau_s$ is the spin relaxation time. The 2nd, and the 4th terms represents the spin precession and spin relaxation respectively.

We can solve $q_{ij}$ from the equation of motion and plug into the continuity equation. Considering that there is only variation along the $z$ direction, we get

$$g\frac{\mu_B}{\hbar}P_y B_z - D\frac{\partial^2 P_x}{\partial z^2} + \frac{P_x}{\tau_s} = 0$$

$$-g\frac{\mu_B}{\hbar}P_x B_z - D\frac{\partial^2 P_y}{\partial z^2} + \frac{P_y}{\tau_s} = 0$$

$$-D\frac{\partial^2 P_z}{\partial z^2} + \frac{P_z}{\tau_s} = 0$$

The third equation and the boundary condition lead to $P_z=0$.

- $P_y$ and the corresponding *longitudinal* charge current

  Combining the first two equations above, one gets

  $$\lambda_s^4 \frac{\partial^4 P_y}{\partial z^4} - 2\lambda_s^2 \frac{\partial^2 P_y}{\partial z^2} + [1 + (\Omega\tau_s)^2]P_y = 0,$$

where $\Omega = g\frac{\mu_B}{\hbar}B_z$, and $\lambda_s^2 = D\tau_s$ is the spin diffusion length. Using the boundary condition of zero spin current at $z=\pm d/2$, i.e.,

$$q_{zy}\left(z = \pm\frac{d}{2}\right) = -D\frac{\partial P_y}{\partial z} + \theta_{SH} q_{x0} = 0,$$

one finds the solution

$$P_y = \frac{\frac{\theta_{SH}}{D}\frac{\lambda_s}{\kappa} q_{x0}}{e^{\frac{\kappa z}{2\lambda_s}} + e^{-\frac{\kappa z}{2\lambda_s}}}\left(e^{\frac{\kappa z}{\lambda_s}} - e^{-\frac{\kappa z}{\lambda_s}}\right)$$

where $q_{x0} = \sigma E_x$, $\kappa = \sqrt{1 - j\Omega\tau_s}$, and $j = \sqrt{-1}$.

The corresponding contribution to the longitudinal charge current is

$$q_x'(z, B_z) = -\theta_{SH} q_{zy} = -\theta_{SH}\left(-D\frac{\partial P_y}{\partial z} + \theta_{SH} q_{x0}\right) = -\theta_{SH}^2 q_{x0}\left(1 - \frac{e^{\frac{\kappa z}{\lambda_s}} + e^{-\frac{\kappa z}{\lambda_s}}}{e^{\frac{\kappa z}{2\lambda_s}} + e^{-\frac{\kappa z}{2\lambda_s}}}\right)$$

Averaging over the whole thickness ($-d/2 \leq z \leq d/2$),

$$\frac{q_x'(B_z)}{q_{x0}} = -\theta_{SH}^2\left(1 - \frac{2\lambda_s}{\kappa d}\frac{e^{\frac{\kappa d}{\lambda_s}} - e^{-\frac{\kappa d}{\lambda_s}}}{e^{\frac{\kappa d}{2\lambda_s}} + e^{-\frac{\kappa d}{2\lambda_s}}}\right)$$

Here we can calculate the magnetoresistance as

$$\frac{\rho_{xx}(B_z)-\rho_{xx0}}{\rho_{xx0}} = \frac{q_x'(0)-q_x'(B_z)}{q_{x0}} = \theta_{SH}^2 \left( \frac{2\lambda_s}{d} \frac{e^{\frac{d}{2\lambda_s}}-e^{-\frac{d}{2\lambda_s}}}{e^{\frac{d}{2\lambda_s}}+e^{-\frac{d}{2\lambda_s}}} - \frac{2\lambda_s}{\kappa d} \frac{e^{\frac{\kappa d}{2\lambda_s}}-e^{-\frac{\kappa d}{2\lambda_s}}}{e^{\frac{\kappa d}{2\lambda_s}}+e^{-\frac{\kappa d}{2\lambda_s}}} \right)$$

$$= \theta_{SH}^2 \frac{\tanh\left(\frac{d}{2\lambda_s}\right)}{\frac{d}{2\lambda_s}} \left(1 - \frac{1}{\kappa}\frac{\tanh\left(\frac{\kappa d}{2\lambda_s}\right)}{\tanh\left(\frac{d}{2\lambda_s}\right)}\right),$$

where $\rho_{xx}$ is the longitudinal resistivity and $\rho_{xx0}$ is the zero-field value of $\rho_{xx}$. Notice that, it is the real part of the formula that is the magnetoresistance.

- $P_x$ and the corresponding *transverse* charge current

From $P_y$, one can derive using $P_x = \frac{D\tau_s \frac{\partial^2 P_y}{\partial z^2}-P_y}{\Omega\tau_s}$ and get $P_x = -jP_y$.

This highlights the relation between $P_x$ and $P_y$ in the spin precession process. The corresponding contribution to the transverse charge current is

$$q_y'(z,B_z) = \theta_{SH} q_{zx} = \theta_{SH}\left(-D\frac{\partial P_x}{\partial z}\right) = \theta_{SH}(jD\frac{\partial P_y}{\partial z}) = -j\theta_{SH}^2 q_{x0}\left(\frac{e^{-\frac{\kappa z}{\lambda_s}}+e^{\frac{\kappa z}{\lambda_s}}}{e^{\frac{\kappa z}{2\lambda_s}}+e^{-\frac{\kappa z}{2\lambda_s}}}\right)$$

Averaging over the whole thickness ($-d/2 \leq z \leq d/2$),

$$\frac{q_y'(B_z)}{q_{x0}} = -j\theta_{SH}^2 \left(\frac{2\lambda_s}{\kappa d} \frac{e^{\frac{\kappa d}{\lambda_s}}-e^{-\frac{\kappa d}{\lambda_s}}}{e^{\frac{\kappa d}{2\lambda_s}}+e^{-\frac{\kappa d}{2\lambda_s}}}\right)$$

We can calculate the Hall effect as

$$\frac{\rho_{xy}(B_z)}{\rho_{xx0}} = \frac{-q_y'(B_z)}{q_{x0}} = \theta_{SH}^2 \left(j\frac{2\lambda_s}{\kappa d} \frac{e^{\frac{\kappa d}{2\lambda_s}}-e^{-\frac{\kappa d}{2\lambda_s}}}{e^{\frac{\kappa d}{2\lambda_s}}+e^{-\frac{\kappa d}{2\lambda_s}}}\right)$$

$$= \theta_{SH}^2 \frac{\tanh\left(\frac{d}{2\lambda_s}\right)}{\frac{d}{2\lambda_s}} \left(j\frac{1}{\kappa}\frac{\tanh\left(\frac{\kappa d}{2\lambda_s}\right)}{\tanh\left(\frac{d}{2\lambda_s}\right)}\right)$$

where $\rho_{xy}$ is the transverse resistivity whose zero-field value is zero. Notice that it is the real part of this formula that describes the Hall effect.

- Combined formula for both longitudinal and transverse effects

The longitudinal (MR, $\frac{\rho_{xx}(B_z)-\rho_{xx0}}{\rho_{xx0}}$) and transverse (Hall, $\frac{\rho_{xy}(B_z)}{\rho_{xx0}}$) effects calculated above can be combined as

$$\frac{\Delta\rho_{SHHE}}{\rho_{L0}} = \theta_{SH}^2 \frac{\tanh\left(\frac{d}{2\lambda_s}\right)}{\frac{d}{2\lambda_s}} \left[1 - \frac{\tanh\left(\frac{\kappa d}{2\lambda_s}\right)}{\kappa \cdot \tanh\left(\frac{d}{2\lambda_s}\right)}\right],$$

where $\rho_{L0}=\rho_{xx0}$, $Real(\Delta\rho_{SHHE}) = Real[\rho_{xx}(B_z) - \rho_{xx0}]$ and $Imag(\Delta\rho_{SHHE}) = Real[\rho_{xy}(B_z)]$, which is the Eq. (1) in the main text. In other words, the real and imaginary parts correspond to the magnetoresistance $\Delta\rho_{L,SHHE}$ and the Hall effect $\rho_{T,SHHE}$ respectively.

### 3.2 Dyakonov's scaling rule

Due to the spin-precession nature of the SHHEs, the effective spin-precession time $\tau_s^*$ is expected to govern the SHHEs. For $d/\lambda_s \to \infty$ (thick film limit), $\tau_s^*$ is limited by the spin relaxation time $\tau_s$, so $\tau_s^* = \tau_s$, where $d$ is the film thickness, as illustrated in Fig. 1(a). For $d/\lambda_s \to 0$ (thin film limit), spin precession occurs over the entire film thickness, so $\tau_s^*$ is the actually the spin diffusion time, i.e., $\tau_s^* = \tau_D = \frac{\left(\frac{d}{2}\right)^2}{D} = \frac{\left(\frac{d}{2}\right)^2}{\frac{\lambda_s^2}{\tau_s}} = \tau_s \left(\frac{d}{2\lambda_s}\right)^2$, where $D$ is the spin diffusion coefficient. The dependence of $\tau_s^*$ on both $\tau_s$ and $d$ can be described using $\frac{1}{\tau_s^*} = \frac{1}{\tau_s} + \frac{1}{\tau_D} = \frac{1}{\tau_s}\left[1 + \left(\frac{2\lambda_s}{d}\right)^2\right]$. The SHHEs are expected to scale roughly with $\tau_s^*$.

In addition, we also note that the maximum SHHEs depends on $\lambda_s$. This can be shown using Eq. (1) as

$$\frac{\Delta\rho_{SHHE}(B_z = \infty)}{\rho_{L0}} = \theta_{SH}^2 \frac{\tanh\left(\frac{d}{2\lambda_s}\right)}{\frac{d}{2\lambda_s}}.$$

Since this is a real function, it can be inferred that $\rho_{L,SHHE}(B_z=\infty) = \theta_{SH}^2 \frac{\tanh\left(\frac{d}{2\lambda_s}\right)}{\frac{d}{2\lambda_s}}$ and $\rho_{T,SHHE}(B_z=\infty) = 0$. Therefore, the normalized SHHEs using $\rho_{L,SHHE}(B_z=\infty)$, i.e.,

$$\frac{\Delta\rho_{SHHE}(B_z)}{\Delta\rho_{L,SHHE}(Bz = \infty)} = 1 - \frac{\tanh\left(\frac{\kappa d}{2\lambda_s}\right)}{\kappa \cdot \tanh\left(\frac{d}{2\lambda_s}\right)}$$

is a universal function that has the maximum value of 1.

The Dyakonov's scaling rule can be visualized in the calculated SHHEs shown Fig. 2. $\Delta\rho_{L,SHHE}$ and $\rho_{T,SHHE}$ are calculated according to Eq. (1) and normalized with the maximum longitudinal effect $\Delta\rho_{L,SHHE}(B_z=\infty)$. The dependence of normalized $\Delta\rho_{L,SHHE}$ and $\rho_{T,SHHE}$ on $\Omega\tau_s^*$ has roughly the same shape for $d/\lambda_s$ over several orders of magnitudes.

Dyakonov's scaling rule is not obvious from the derivation of SHHEs because it's not a strict rule. On the other hand, it is critical to consider when one tries to extract spin-transport parameters from the field-dependence of SHHEs experimentally. Experimentally, one measures the magnetic-field ($B$) dependence of SHHE signals. The fine differences between curves of different $d/\lambda_s$ in Fig. 2 are, unfortunately not always resolvable considering the experimental uncertainty. Therefore, experimentally, from the shape of the field-dependence of SHHEs, one can only derive $\Omega\tau_s^*$ directly, according to the scaling rule. On the other hand, for the same $\tau_s^*$, there are multiple sets of ($\tau_s$, $\lambda_s$) to satisfy $\frac{1}{\tau_s^*} = \frac{1}{\tau_s}\left[1 + \left(\frac{2\lambda_s}{d}\right)^2\right]$. To determine the spin transport parameters unambiguously, we have resorted to the Pt thickness dependence of SHHE signals to derive $\lambda_s$ first and find $\tau_s$ from $\tau_s^*$ based on the known $\lambda_s$, as explained in in **Section 3.3** below.

### 3.3 Thickness dependence of low-field spin Hall Hanle effect (SHHE)

In principle, fitting the field-dependence of SHHE signals based on Eq. (1) allows one to derive spin transport parameters, such as $\theta_{SH}$ and $\lambda_s$, in heavy metals. However, the values of $\theta_{SH}$ and $\lambda_s$ derived in such a way are not unique, simply because both quantities are entangled with each other in Eq. (1). As illustrated by the scaling rule in Figure 2, a small $\lambda_s$ value (together with a large $\theta_{SH}$ value) can generate a fitting comparable to the one generated by a large $\lambda_s$ value (together with a small $\theta_{SH}$ value). To extract a unique set of $\theta_{SH}$ and $\lambda_s$, we must have knowledge of either one of the two quantities, then derive the other by fittings based on Eq. (1). Previous work [1] on SHHE has cited the value of $\theta_{SH}$ from spin pumping experiments, and derived $\lambda_s$ and $D$ from the fitting. Here we want to demonstrate that the scaling rule can be circumvented using low-field approximation and all the spin transport parameters can be derived from the thickness dependence and the field dependence of $\Delta\rho_L/\rho_{L0}$ and $\rho_T/\rho_{L0}$.

A thorough analysis [1] of Eq. (1) reveals that, in the low field regime ($\Omega\tau_s \ll 1$), $\rho_T/\rho_{L0}$ depends linearly on the magnetic field ($\Omega = g\mu_B B/\hbar$, where $B$ is the field along the out-of-plane direction),

$$\frac{\rho_T}{\rho_{L0}} \approx \theta_{SH}^2 \frac{\frac{\lambda_s}{d}\sinh\left(\frac{d}{\lambda_s}\right) - 1}{1 + \cosh\left(\frac{d}{\lambda_s}\right)} \Omega\tau_s,$$

while $\Delta\rho_L/\rho_{L0}$ depends quadratically on the magnetic field:

$$\frac{\Delta\rho_L}{\rho_{L0}} \approx \frac{\theta_{SH}^2}{8}\left[\frac{6\lambda_s}{d}\tanh\left(\frac{d}{2\lambda_s}\right) - \frac{3 + \frac{d}{\lambda_s}\tanh\left(\frac{d}{2\lambda_s}\right)}{\cosh\left(\frac{d}{2\lambda_s}\right)}\right](\Omega\tau_s)^2.$$

Now, one can consider the curvature of the field-dependence of $\Delta\rho_L/\rho_{L0}$ in the low field regime:

$$C\left(\frac{d}{\lambda_s}\right) \equiv \frac{\partial^2\left(\frac{\Delta\rho_L}{\rho_{L0}}\right)}{\partial B^2} \approx \frac{\theta_{SH}^2}{8}\left[\frac{6\lambda_s}{d}\tanh\left(\frac{d}{2\lambda_s}\right) - \frac{3 + \frac{d}{\lambda_s}\tanh\left(\frac{d}{2\lambda_s}\right)}{\cosh\left(\frac{d}{2\lambda_s}\right)}\right]\left(\frac{g\mu_B}{\hbar}\tau_s\right)^2.$$

The shape of $C(d/\lambda_s)$ is determined solely by the factor $\left[\frac{6\lambda_s}{d}\tanh\left(\frac{d}{2\lambda_s}\right) - \frac{3+\frac{d}{\lambda_s}\tanh\left(\frac{d}{2\lambda_s}\right)}{\cosh\left(\frac{d}{2\lambda_s}\right)}\right]$, while $\theta_{SH}$ and $\tau_s$ only affects the magnitude. Numeric simulation shows that $C$ reaches maximum at $d/\lambda_s \approx 4.56$, as shown in Fig. S3(a).

Similar, one can consider the slope of the field dependence of $\rho_T/\rho_{L0}$ in the low-field regime:

$$S\left(\frac{d}{\lambda_S}\right) \equiv \frac{\partial\left(\frac{\Delta\rho_L}{\rho_{L0}}\right)}{\partial B} \approx \theta_{SH}^2 \frac{\frac{\lambda_s}{d}\sinh\left(\frac{d}{\lambda_s}\right) - 1}{1+\cosh\left(\frac{d}{\lambda_s}\right)}\left(\frac{g\mu_B}{\hbar}\tau_s\right).$$

Again, the shape of $S(d/\lambda_s)$ is determined solely by the factor $\frac{\frac{\lambda_S}{d}\sinh\left(\frac{d}{\lambda_S}\right)-1}{1+\cosh\left(\frac{d}{\lambda_S}\right)}$, while $\theta_{SH}$ and $\tau_s$ only affects the magnitude. Numeric simulation shows that $S$ reaches maximum at $d/\lambda_s \approx 3.28$, as shown in Fig. S3(b).

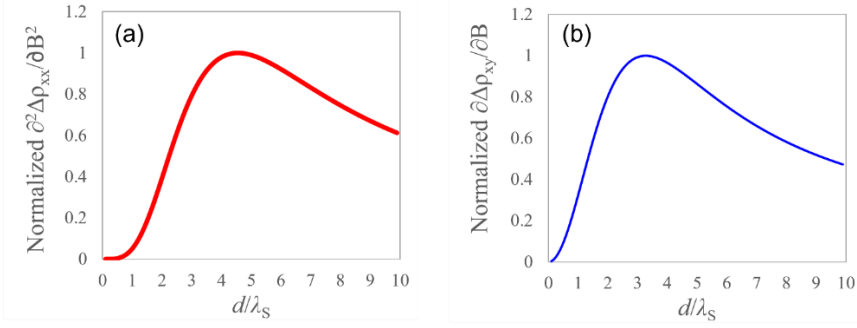

**Fig. S3** Numerical simulation of the Pt thickness dependence of low-field (a) $\Delta\rho_L/\rho_{L0}$ curvature and (b) $\Delta\rho_T/\rho_{L0}$ slope based on Eq. (1) in the main text.

Therefore, by measuring the thickness dependence of $S(d/\lambda_s)$ and $C(d/\lambda_s)$, one can determine $\lambda_s$ by matching the experimental shape and the simulation in Fig. S3. In this work, we derive $\lambda_s \approx 1.63$ nm based on the simultaneous fittings of the thickness dependence of $\Delta\rho_L/\rho_{L0}$ and $\rho_T/\rho_{L0}$ at low fields. Once we have a good estimation of $\lambda_s$, we then derive $\theta_{SH}$ and $D$ (or $\tau_s$) by fitting the field dependence of $\Delta\rho_L/\rho_{L0}$ and $\Delta\rho_T/\rho_{L0}$, which are summarized in Table 1.

## S4. Fitting of the field dependence of $\Delta\rho_L/\rho_{L0}$ and $\Delta\rho_T/\rho_{L0}$

As discussed in the main text, ordinary MR contributes to the total longitudinal MR measured in all axes, the SHHE-induced longitudinal component $\Delta\rho_L/\rho_{L0}$ is calculated as $[\Delta\rho_L(B_z) - \Delta\rho_L(B_y)]/\rho_{L0}$. Meanwhile, ordinary Hall effect contributes to the total transverse signal measured in z axis, giving a linear background to $\rho_T/\rho_{L0}$. During fitting, a same set of $\theta_{SH}$, $\lambda_s$ and $D$ are used to fit both $\Delta\rho_L/\rho_{L0}$ and $\rho_T/\rho_{L0}$.

In Fig. S4, the top 3 panels are fits of $\Delta\rho_L/\rho_{L0}$ for samples with Pt thickness of 3.8 nm, 5.2 nm, and 6.3 nm, respectively; the bottom 3 plots are fittings of $\rho_T/\rho_{L0}$ correspondingly. We can see clearly that the longitudinal MR signals start to deviate from quadratic shape when $B_z$ is above 4 T and gradually saturate at higher fields. This is because faster spin precession causes a wider distribution of projections onto $\hat{y}$ axis from reflected spin current $\vec{J}_{S,R}$, which leads to more cancellation of $\vec{J}_{C,R}$ projections along $\hat{x}$ axis and thus less longitudinal MR change. Meanwhile, the transverse signals contributed by SHHE also gradually saturate when strong-precession condition is met, leaving a linear slope at high fields due to ordinary Hall effect (OHE). The SHHE components and the OHE components of transverse signals are indicated by blue and red solid curves, respectively, in Figure S4.

We want to emphasize that the non-linear transverse signal is a hallmark feature that differentiates SHHE from OHE and reveals the influence of spin precession in modifying the spin accumulation at the sample boundaries. Similar non-linear transverse Hall effect has been observed in Pt/yttrium iron garnet system but has been broadly regarded as anomalous Hall effect with polarity changing from positive at low temperatures to negative at high temperatures [2, 3]. Since SHHE is a ubiquitous phenomenon in heavy metals at high fields, its contribution to longitudinal magnetoresistance and transverse Hall effect should be included in the interpretation of complex magneto-transport phenomena from now on.

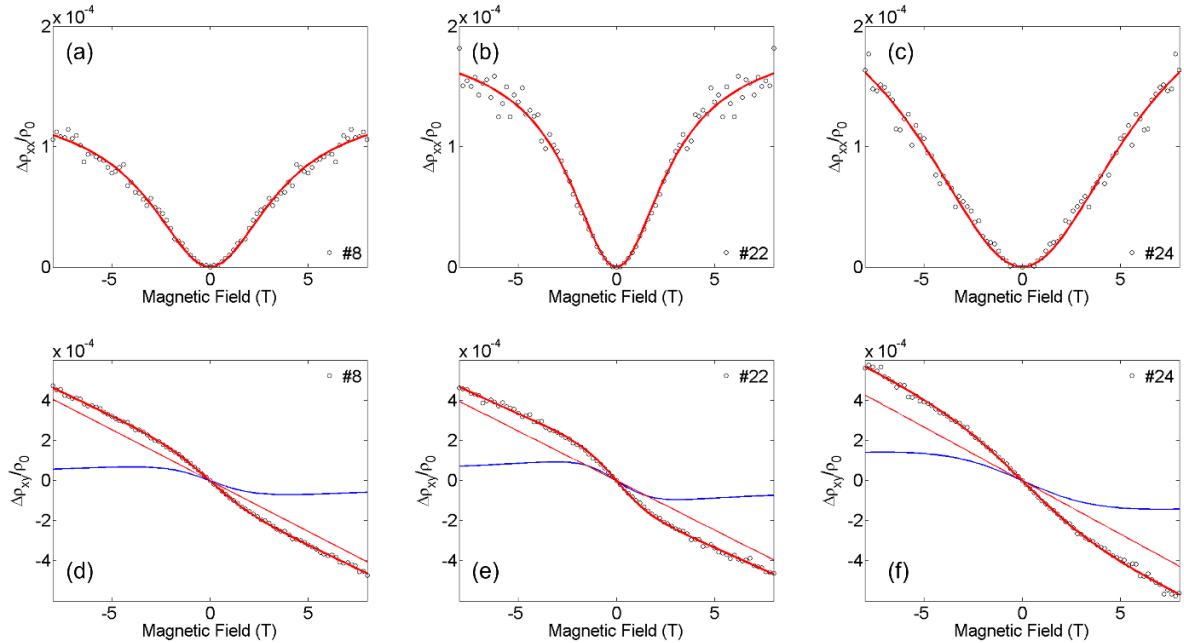

**Fig. S4** Fitting of field dependence of (a-c) longitudinal and (d-f) transverse signals from three Pt/Al$_2$O$_3$ samples with Pt thickness of 3.8 nm (#8), 5.2 nm (#22), and 6.3 nm (#24), respectively. Hollow circle symbols represent experimental data and solid curves represent fittings. The SHHE components and the OHE components of transverse signals are represented by blue and red solid curves, respectively.